\documentclass[10pt,aps,prl,twocolumn,groupedaddress,showkeys,floatfix]{revtex4-2}
\usepackage{amsmath}
\usepackage{graphicx}
\usepackage{ragged2e}
\usepackage[colorlinks=true,linkcolor=blue,urlcolor=blue,citecolor=blue,anchorcolor=blue]{hyperref}
\DeclareGraphicsExtensions{.pdf,.eps,.png,.jpg,.mps}
\usepackage{epstopdf}
\usepackage{threeparttable}
\usepackage{color}
\usepackage[T2A]{fontenc}
\usepackage[utf8]{inputenc}
\usepackage{float}
\usepackage{tabularx}
\usepackage{array}
\usepackage[percent]{overpic}
\usepackage{amsmath}
\begin{document}
%TC:ignore
\title{Alloy engineering of excitonic properties in TMD monolayers}

\author{
Eirini~Katsipoulaki$^{1}$,
Adlen~Smiri$^{2}$,
Panagiotis~Spiliotakis$^{1,3}$,
Konstantinos~Mourzidis$^{2}$,
Danae~Katrisioti$^{1,3}$,
Takashi~Taniguchi$^{4}$,
Kenji~Watanabe$^{5}$,
Georgios~Kopidakis$^{1,3}$,
Zden\v{e}k~Sofer$^{6}$,
Gang~Wang$^{7}$,
Emmanuel~Stratakis$^{1}$,
George~Kioseoglou$^{1,3}$,
Iann~C.~Gerber$^{2,\dagger}$,
Xavier~Marie$^{2,8,\#}$,
and Ioannis~Paradisanos$^{1,3,*}$
}

\address{$^{1}$Institute of Electronic Structure and Laser, Foundation for Research and Technology-Hellas, Heraklion 71110, Greece}

\address{$^{2}$Univ Toulouse, INSA, CNRS, LPCNO, Toulouse, France}

\address{$^{3}$Department of Materials Science and Engineering, University of Crete, Heraklion 70013, Greece}

\address{$^{4}$Research Center for Materials Nanoarchitectonics, National Institute for Materials Science, 1-1 Namiki, Tsukuba 305-0044, Japan}

\address{$^{5}$Research Center for Electronic and Optical Materials, National Institute for Materials Science, 1-1 Namiki, Tsukuba 305-0044, Japan}

\address{$^{6}$Department of Inorganic Chemistry, University of Chemistry and Technology Prague, Technick\'a 5, 166 28 Prague 6, Czech Republic}

\address{$^{7}$Key Laboratory of Advanced Optoelectronic Quantum Architecture and Measurement, Ministry of Education, School of Physics and Beijing Key Laboratory of Nanophotonics and Ultrafine Optoelectronic Systems, Beijing Institute of Technology, Beijing, China}

\address{$^{8}$Institut Universitaire de France, 75231 Paris, France}

\address{$^{\dagger}$ igerber@insa-toulouse.fr,
$^{\#}$ marie@insa-toulouse.fr,
$^{*}$ iparad@iesl.forth.gr}


\begin{abstract}
\section{\label{Abstract}Abstract}
We investigate monolayer MoS$_{2x}$Se$_{2(1-x)}$ alloys across the full composition range using optical spectroscopy. We demonstrate continuous tuning of the optical gap over $\sim$0.35 eV, accompanied by a systematic reduction of the B--A exciton splitting, in agreement with density functional theory calculations. Temperature-dependent measurements reveal a progressive increase of the average phonon energy from Se-rich to S-rich alloys that follows a simple reduced-mass scaling model. Polarization-resolved spectroscopy further shows a monotonic increase of the circular polarization from nearly zero in MoSe$_2$ to $\sim$15\% in MoS$_2$ at 78 K. The observed evolution of the polarization is attributed to alloy-induced modifications of the electronic structure that modify bright--dark exciton mixing and the associated valley depolarization. These findings establish alloy engineering as an effective strategy for controlling excitonic properties in TMD monolayers.

\end{abstract}

\maketitle
%TC:endignore
\section{\label{Intro}Introduction}

Two-dimensional (2D) transition-metal dichalcogenides (TMDs) represent a rich platform of atomically thin semiconductors combining direct bandgaps~\cite{splendiani2010emerging,mak2010atomically} with strong spin--orbit coupling (SOC)~\cite{zeng2013optical,mak2012control,ren2023measurement} in the visible to near-infrared spectral range~\cite{yu20172d}. In monolayer TMDs, time-reversal symmetry, together with broken inversion symmetry and SOC, gives rise to coupled spin and valley degrees of freedom at the inequivalent K$^+$ and K$^-$ valleys of the Brillouin zone~\cite{liu2015electronic,xiao2012coupled}. As a consequence, circularly polarized light can selectively induce transitions in the K$^+$ and K$^-$ valleys, allowing optical initialization and readout of valley polarization~\cite{xiao2012coupled}. These properties have stimulated extensive research toward applications in optoelectronics~\cite{wang2012electronics}, valleytronics~\cite{schaibley2016valleytronics,seyler2026valleytronics2dmaterialsroadmap}, and quantum photonics~\cite{paralikis2025tunable}.\\
\indent Unlike external perturbations such as strain~\cite{kourmoulakis2023biaxial, michail2023tuning, PhysRevB.88.121301, feierabend2017impact, maniadaki2016strain} or doping~\cite{katsipoulaki2023electron, katsipoulaki2025spin,yue2013functionalization,coelho2018post, wang2017progress, yang2022relaxation}, semiconductor alloying provides a stable and synthesis-controlled route to tailor material properties, with compatibility for large-area growth techniques. In conventional semiconductors, alloy systems such as AlGaAs, InGaAs, and InGaN have been crucial for the development of light-emitting diodes (LEDs)~\cite{nakamura1994candela}, laser diodes~\cite{razeghi1994high}, and optical communication technologies~\cite{lin2021ingan}, while II--VI alloys such as CdZnTe and HgCdTe play a key role in infrared detection~\cite{rogalski2005hgcdte}. Similarly, SiGe alloys have been exploited in high-speed electronics and photonic integration within silicon platforms~\cite{fadaly2020direct}. These concepts naturally extend to 2D-TMDs, where alloying provides a convenient tool to tailor their electronic, excitonic, and spin–valley properties. Alloying can be achieved either through chalcogen substitution, as in MoS$_{2x}$Se$_{2(1-x)}$~\cite{ma2014postgrowth, gong2014band,li2014growth,feng2014growth}, or through transition-metal substitution, such as in Mo$_{1-x}$W$_x$-based compounds~\cite{zhang2014two, chen2013tunable,wang2015spin,zheng2015monolayers}, both of which have been shown to modify key optoelectronic properties~\cite{chen2013tunable,komsa2012two,susarla2017quaternary,liu2020room, xie2015two}.\\
\indent Despite these advances, most studies have primarily focused on bandgap engineering and excitonic transition energies~\cite{zhang2014two,chen2013tunable,feng2015growth,feng2014growth,fu2015synthesis,komsa2012two,kang2013monolayer}. In contrast, the systematic evolution of vibrational and spin--valley properties across the full alloy composition range remains comparatively less explored. In particular, the composition dependence of average phonon energies, spin--orbit-related excitonic splittings, and valley polarization has not been investigated within a unified experimental framework. Optical spectroscopy can therefore provide valuable insight into how alloy composition modifies the electronic structure and optical response of TMD alloys.\\
\begin{figure*}[tb]
\centering
\includegraphics[width=\textwidth]{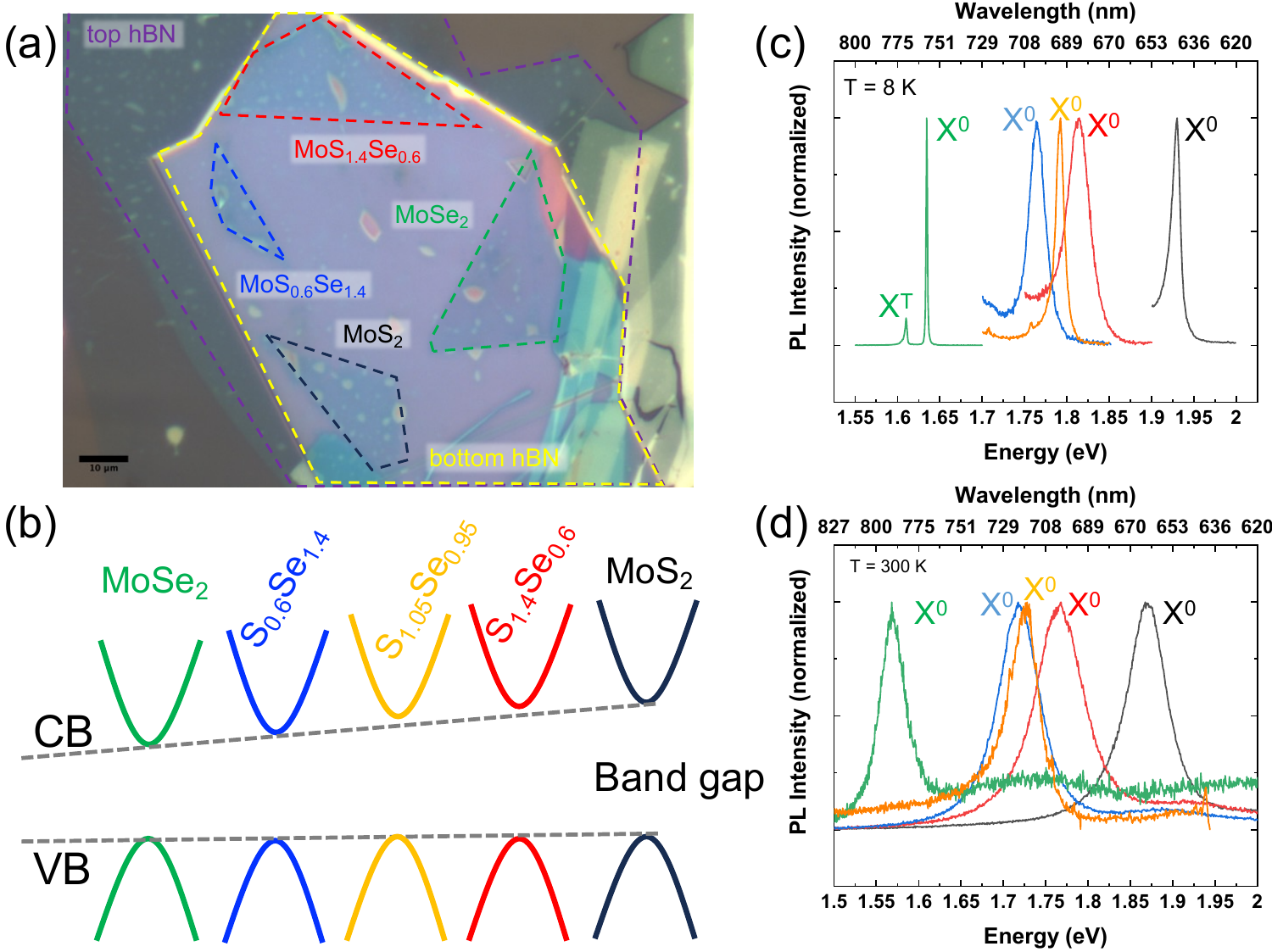}
\caption{(a) Optical microscope image of encapsulated MoSe$_2$ (green dashed line), MoS$_{0.6}$Se$_{1.4}$ (blue dashed line), MoS$_{1.4}$Se$_{0.6}$ (red dashed line), and MoS$_2$ (black dashed line) monolayers. (b) Schematic illustration of the bandgap evolution with different chalcogen composition in MoS$_{2x}$Se$_{2(1-x)}$ alloys. (c) Normalized PL spectra of MoSe$_2$ (green), MoS$_{0.6}$Se$_{1.4}$ (blue), MoS$_{1.05}$Se$_{0.95}$ (orange), MoS$_{1.4}$Se$_{0.6}$ (red), and MoS$_2$ (black) measured at 8 K and (d) at 300 K.}
\label{fig:fig1}
\end{figure*}
\indent In this work, we combine temperature- and polarization-dependent optical spectroscopy with density functional theory (DFT) calculations to investigate monolayer MoS$_{2x}$Se$_{2(1-x)}$ alloys encapsulated in hexagonal boron nitride (hBN). We systematically examine the evolution of the optical bandgap, B--A exciton splitting, average phonon energies, and exciton circular polarization across the full alloy composition range. The excitonic transition energies exhibit nearly Vegard-like behavior with negligible bowing parameter, while the experimentally observed evolution of the B--A exciton splitting is in close agreement with DFT-calculated spin--orbit splittings. Temperature-dependent spectroscopy further reveals a systematic increase of the average phonon energy from $\sim$17~meV in Se-rich to $\sim$22~meV in S-rich alloys, which is well described by a reduced-mass scaling model. Finally, polarization-resolved measurements reveal a monotonic increase of the exciton circular polarization with increasing sulfur content. DFT calculations provide insight into this behavior through alloy-induced modifications of the spin-split conduction bands. These findings demonstrate how chalcogen alloying offers controlled tuning of excitonic, vibrational, and spin--valley properties in two-dimensional semiconductors.
\begin{figure*}[tb]
\centering
\includegraphics[width=\textwidth]{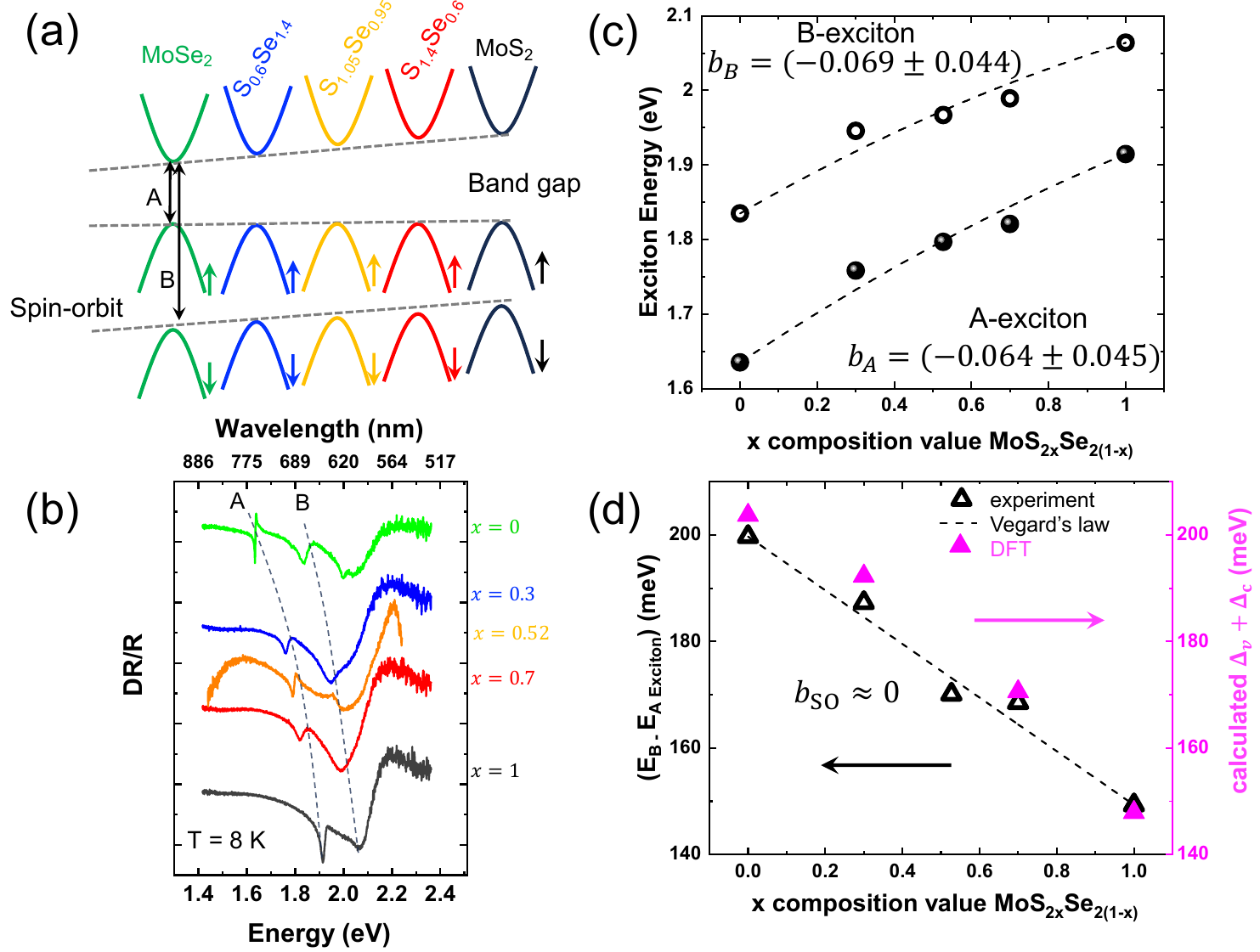}
\caption{(a) Schematic illustration of the evolution of the spin–orbit splitting of the valence bands with alloy composition, showing a decrease of the A–B splitting from MoSe$_2$ to MoS$_2$. While the illustration emphasizes the dominant valence-band spin--orbit splitting, the experimentally measured B--A exciton energy separation also contains a smaller contribution from the conduction-band splitting, which is not shown here.
(b) Differential reflectivity spectra at 8 K for MoS$_{2x}$Se$_{2(1-x)}$ alloys with $x=0, 0.3, 0.52, 0.7, 1$, highlighting the A and B excitonic resonances.
(c) Composition dependence of the A and B exciton resonances extracted from PL spectra at 78 K. Dashed lines are fits using Eq.(\ref{eq:bandgap}), yielding small bowing parameters ($b_A$, $b_B$).
(d) Energy difference between the B and A excitons measured experimentally (black triangles) as a function of alloy composition, together with the DFT-calculated energy splitting of the spin-aligned band-edge electronic states (magenta filled triangles). The data show an approximately linear decrease from MoSe$_2$ to MoS$_2$, consistent with a negligible bowing parameter ($b_{\mathrm{SO}} \approx 0$).}
\label{fig:soc}
\end{figure*}

\section{\label{results}Results and Discussion}
This study focuses on monolayers of MoS$_{2x}$Se$_{2(1-x)}$ alloys with compositions ranging from x=0 to x=1, specifically MoSe$_2$, MoS$_{0.6}$Se$_{1.4}$, MoS$_{1.05}$Se$_{0.95}$, MoS$_{1.4}$Se$_{0.6}$, MoS$_2$.
An optical microscope image of the sample is shown in Fig.~\ref{fig:fig1}(a). Dashed lines indicate the different monolayer regions: MoSe$_2$ (green), MoS$_{0.6}$Se$_{1.4}$ (blue), MoS$_{1.4}$Se$_{0.6}$ (red), and MoS$_2$ (black). The MoS$_{1.05}$Se$_{0.95}$ monolayer was measured on a different sample and is therefore not shown in Fig.~\ref{fig:fig1}(a). The evolution of the bandgap with alloying is illustrated in Fig.~\ref{fig:fig1}(b) for the five different MoS$_{2x}$Se$_{2(1-x)}$ monolayer samples studied here. Progressive substitution of selenium (Se) by sulfur (S) results in a monotonic increase of the bandgap from MoSe$_2$ to MoS$_2$. The corresponding excitonic ground-state transition (optical gap) is therefore anticipated to follow the same compositional trend, given the similar exciton binding energies reported for encapsulated MoS$_2$ ($\sim221$ meV) and MoSe$_2$ ($\sim231$ meV)~\cite{goryca2019revealing}. This behavior is confirmed experimentally in the composition-dependent PL spectra shown in Fig.~\ref{fig:fig1}(c), where the neutral exciton (X$^0$) emission exhibits a continuous blueshift from 1.63 eV in MoSe$_2$ to 1.93 eV in MoS$_2$ at low temperature (8 K). This monotonic blueshift of the transition energy with alloy composition is also observed at room temperature  (Fig.~\ref{fig:fig1}(d)), even though thermal broadening leads to substantially wider emission features compared to 8 K. These PL results demonstrate that chalcogen alloying allows the optical gap to be continuously tuned~\cite{gong2014band} over a wide spectral range of $\sim$130 nm (0.35 eV) across the Mo-based TMD alloy series.
\begin{figure*}[tb]
\centering
\includegraphics[width=\textwidth]{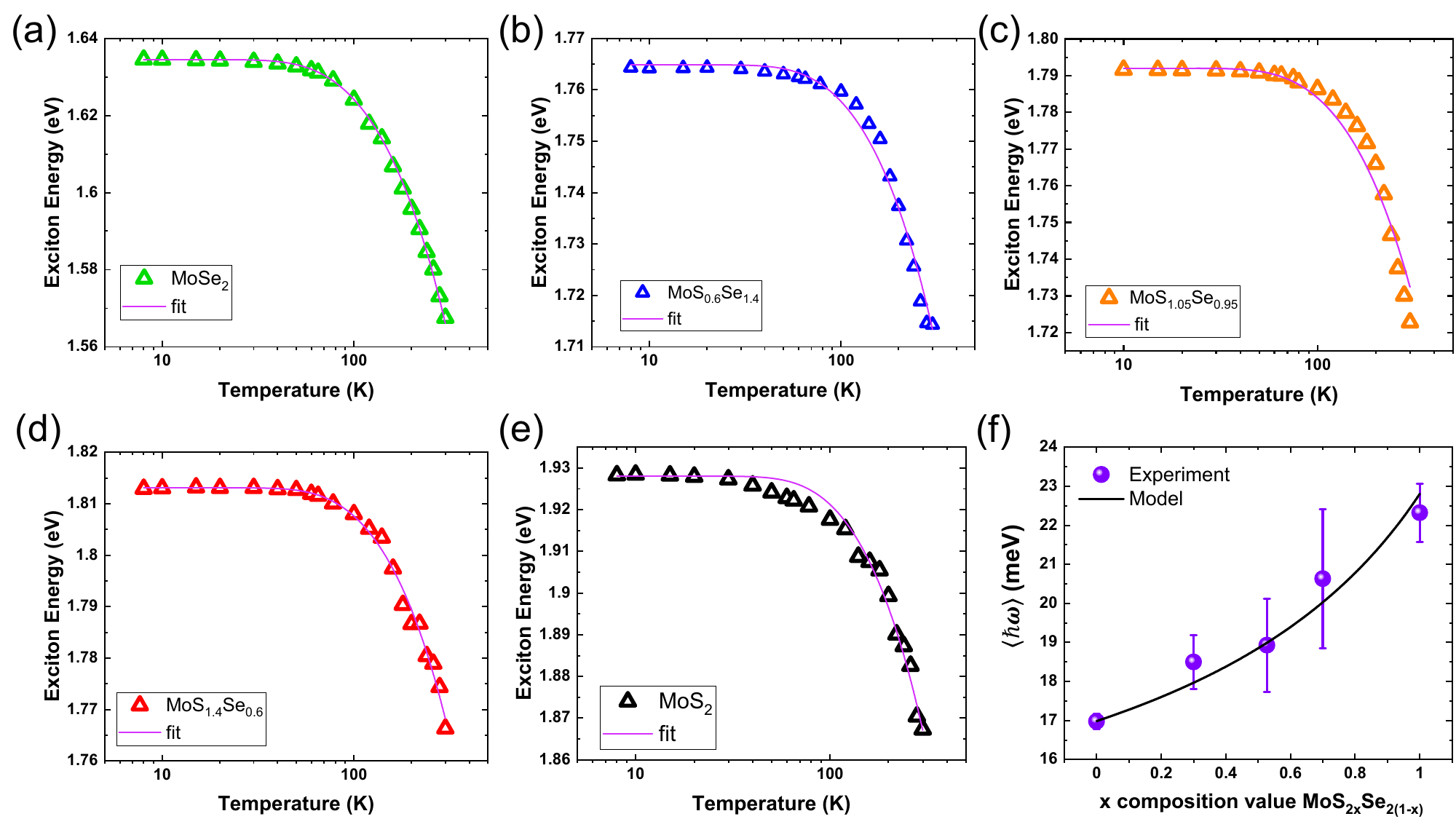}
\caption{(a) Temperature dependence of A-exciton transition energy for (a) MoSe$_2$, (b) MoS$_{0.6}$Se$_{1.4}$, (c) MoS$_{1.05}$Se$_{0.95}$, (d) MoS$_{1.4}$Se$_{0.6}$, (e) MoS$_2$. Triangles represent experimental data extracted from PL measurements, and solid lines are fits using Eq.(\ref{eq:odonel}). (f) Average phonon energy $\langle \hbar\omega \rangle$ extracted from the temperature-dependent PL analysis as a function of alloy composition $x$ in $\mathrm{MoS}_{2x}\mathrm{Se}_{2(1-x)}$ monolayers. Symbols represent experimental values, while the solid line shows a fit based on a reduced-mass scaling model.
}
\label{fig:phonon}
\end{figure*}
\indent Alloying is also expected to modify the magnitude of the spin–orbit splitting, as MoS$_2$ and MoSe$_2$ exhibit different spin–orbit coupling strengths~\cite{ramasubramaniam2012large}, primarily due to the enhanced contribution of the heavier chalcogen atom in MoSe$_2$~\cite{kosmider2013large, kormanyos2015k}. In optical spectroscopy experiments, this splitting is not accessed directly, but is reflected in the energy separation between the A and B exciton transitions. Fig.~\ref{fig:soc}(a) schematically illustrates the expected trends: while the bandgap increases with sulfur content, the spin--orbit splitting of the band-edges decreases from MoSe$_2$ to MoS$_2$. Consequently, the A-B exciton separation is predicted to decrease monotonically with increasing sulfur concentration in the case of MoS$_{2x}$Se$_{2(1-x)}$ alloys. This behavior is confirmed by photoluminescence excitation (PLE) spectroscopy (Fig. S1) and by differential reflectivity measurements at 8 K (Fig.~\ref{fig:soc}(b)), where the A and B excitonic resonances are clearly resolved across all compositions. The gray dashed lines serve as guides to the eye, highlighting the evolution of the excitonic transitions with chalcogen composition. Notably, the A–B splitting decreases systematically from MoSe$_2$ to MoS$_2$, with intermediate alloy compositions exhibiting values between those of the two pristine compounds.\\
\indent To quantify the compositional dependence, we analyze the exciton energy for A and B excitonic transitions as a function of alloy composition in Fig.~\ref{fig:soc}(c). The data were extracted from PL spectra at 8 K (Fig. S2, supplementary material). Similar behavior is observed in reflectivity and PLE experiments(Fig. S1). The compositional dependence of the exciton transition energies can be understood within the framework of Vegard’s law, which states that a material’s lattice parameter in a solid solution varies approximately linearly with composition~\cite{vegard1921constitution,denton1991vegard}. In semiconductors, the energy bandgap 
\textit{E$_g$} is approximately a linear function of the lattice parameter. Therefore, if the lattice parameter of a semiconductor follows Vegard’s law, a linear relationship between the bandgap and the composition is expected. Deviations from linear Vegard interpolation are quantified by the bowing parameter \textit{b}, which measures the curvature of the compositional dependence. Physically, the bowing parameter reflects the effects of lattice-constant and atomic-size mismatch between the alloy constituents, which lead to departures from linear behavior~\cite{tongay2014two}. For similar exciton binding energies, the empirical expression of the optical gap energy, $E_g^{\mathrm{opt}}$, as a function of the composition $x$, is given by:
\begin{equation}
E_g^{\mathrm{opt}}(x) =
xE_g^{\mathrm{opt}}(\mathrm{MoS}_2)
+ (1-x)E_g^{\mathrm{opt}}(\mathrm{MoSe}_2)
- bx(1-x)
\label{eq:bandgap}
\end{equation}
where \textit{x} is the sulfur fraction and \textit{b} is the bowing parameter that quantifies deviations from linear interpolation. This relation was used to fit the experimental A- and B-exciton energies in Fig.~\ref{fig:soc}(c), with the resulting fittings shown as black dashed lines. The extracted bowing parameters $b_A$=(-0.064±0.045)eV and $b_B$=(-0.069±0.044)eV, are small~\cite{feng2015growth,bendavid2022strain} indicating negligible deviations from linear compositional dependence. We further note that the DFT-calculated bandgap energies exhibit similarly small bowing parameters (Fig.~S3), in agreement with the experimentally observed behavior of the A and B excitons. This near-linear behavior suggests that alloying introduces limited perturbations to the electronic structure, consistent with the similar band-edge characteristics of MoS$_2$ and MoSe$_2$. In contrast, a large bowing parameter would reflect strong alloy-induced effects, such as local structural distortions, enhanced disorder, and significant modifications of the band-edge states, leading to pronounced nonlinearity in the excitonic transition energies~\cite{yin2008origin}.\\
\indent The dependence of the energy difference between B and A excitons on alloy composition (black triangles), extracted from the PL spectra at 8 K (Fig. S2, supplementary material), is summarized in Fig.~\ref{fig:soc}(d). Similar trends are observed in PLE spectroscopy (Fig. S1) and reflectivity experiments. The splitting decreases monotonically from $\sim$200 meV in MoSe$_2$ to $\sim$150 meV in MoS$_2$, reflecting the reduction of the spin--orbit splitting of the band-edges with increasing sulfur content, in agreement with previously reported values in the literature~\cite{zhu2011giant,ciesiolkiewicz2026sensitivity}. The linear variation of the energy difference between B and A excitons with alloy composition is well described by Vegard’s law (black dashed line in Fig.~\ref{fig:soc}(d)), indicating a vanishing bowing parameter, $b_{\mathrm{SO}}$. It is important to note that, for monolayers in which the lowest conduction band and the upper valence band are spin aligned, the energy difference between the B and A excitons is given by:
\begin{equation}
E_B - E_A = \Delta_v + \Delta_c - (E_b^B - E_b^A)
\end{equation}
Here, $\Delta_v$ and $\Delta_c$ denote the spin--orbit splittings of the valence and conduction bands, respectively, while $E_b^A$ and $E_b^B$ are the binding energies of the A and B excitons. Therefore, the experimentally measured B--A exciton splitting is not determined solely by the valence-band spin--orbit splitting, but also contains contributions from the smaller conduction-band splitting~\cite{ren2023measurement} and from differences in the exciton binding energies. Based on established values of the valence- and conduction-band effective masses at the $K$ point~\cite{kormanyos2015k}, we infer that the reduced masses of the A and B excitons remain very similar in MoS$_2$- and MoSe$_2$-based systems, suggesting that the corresponding difference in binding energies is not expected to be very large. Additional corrections, such as electron--hole exchange interactions, may also contribute but are generally expected to remain comparatively small. With these considerations in mind, the experimentally measured B--A exciton splitting can still be qualitatively compared with the DFT-calculated energy splitting of the spin-aligned band-edge states shown in Fig.~\ref{fig:soc}(d) (magenta filled triangles), where each point corresponds to $\Delta_v+\Delta_c$ for the corresponding alloy composition. The calculated splittings are found to follow closely the experimentally observed compositional trend and remain relatively close in magnitude to the measured B--A exciton energy separation (black triangles). We note, however, that the calculated splittings for the intermediate alloy compositions are sensitive to the specific Se/S atomic arrangement considered in the simulations. Since the DFT calculations were performed using a single random atomic configuration for each composition, a more quantitative estimation of the associated uncertainty would ideally require averaging over multiple alloy configurations.\\
\indent Temperature-dependent PL measurements of the exciton transition energy provide insight into the average phonon energies. The temperature dependence of the A-exciton transition energy for the different alloy compositions is shown in Fig.~\ref{fig:phonon}(a)-(e). The A-exciton energy was extracted through Lorentzian fitting of the temperature dependent PL spectra (Fig. S4 of the supplementary material) in order to minimize possible contributions from trion states to the neutral exciton energy. The PL temperature dependence was analyzed using the following empirical expression for the bandgap energy~\cite{o1991temperature}:
\begin{equation}
E_g(T) = E_g(0) - S\langle \hbar \omega \rangle 
\left[ \coth\left(\frac{\langle \hbar \omega \rangle}{2k_B T}\right) - 1 \right]
\label{eq:odonel}
\end{equation}
where E$_g$(0) is the bandgap at zero temperature, \textit{S} is the dimensionless exciton-phonon coupling constant, and $\langle \hbar \omega \rangle$ represents the average phonon energy. The model (solid red lines in Fig.~\ref{fig:phonon}(a)–(e)) provides a reasonable description of the experimental data (triangles in Fig.~\ref{fig:phonon}(a)–(e)), allowing the extraction of $\langle \hbar \omega \rangle$ for each alloy composition.
\begin{figure*}[tb]
\centering
\includegraphics[width=\textwidth]{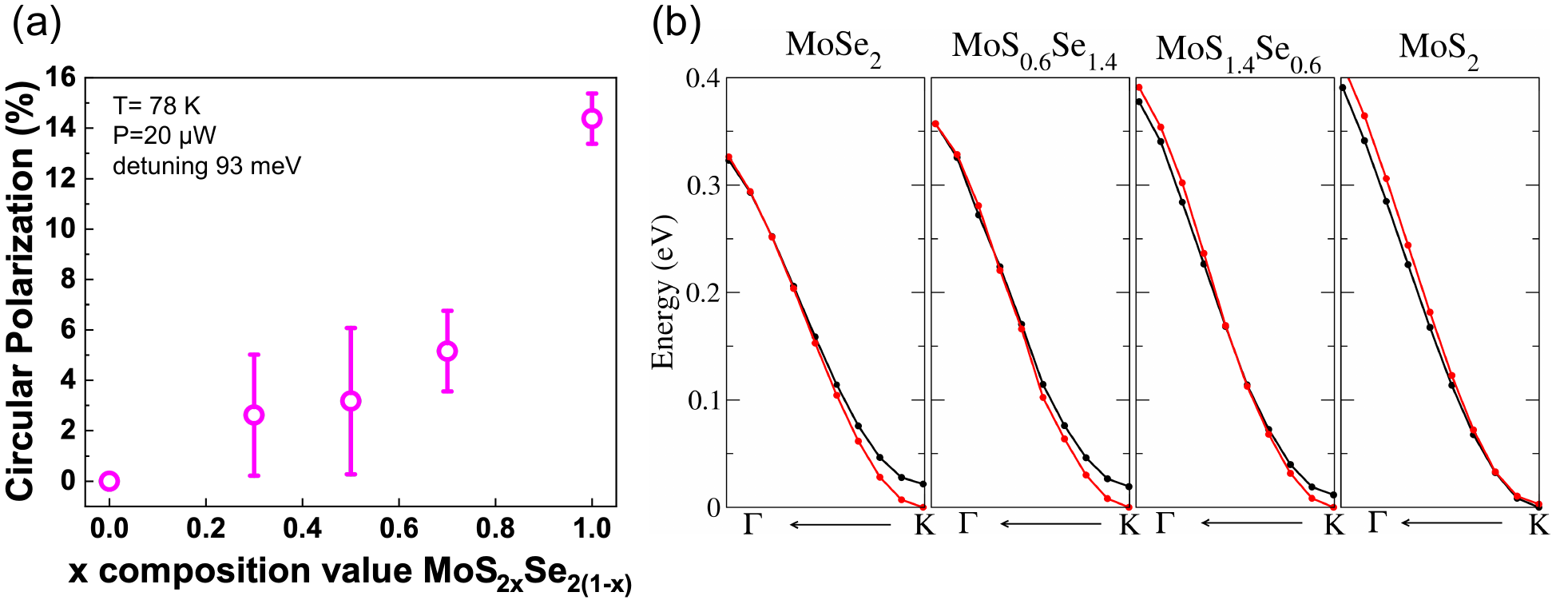}
\caption{
(a) Experimentally measured exciton circular polarization as a function of alloy composition, showing a monotonic increase with increasing sulfur content. (b) DFT-calculated evolution of the two lowest conduction bands near the $K$ point for different alloy compositions: MoSe$_2$, MoS$_{0.6}$Se$_{1.4}$, MoS$_{1.4}$Se$_{0.6}$, and MoS$_2$. The spin splitting near the $K$ valley decreases with increasing sulfur content, reflecting the progressive modification of the electronic structure induced by alloying.
}
\label{fig:polarization}
\end{figure*}
The extracted average phonon energies are summarized in Fig.~\ref{fig:phonon}(f) and exhibit a clear monotonic increase from $\sim$17 meV in MoSe$_2$~\cite{ross2013electrical,kioseoglou2016optical} to $\sim$22.5 meV in MoS$_2$~\cite{cadiz2017excitonic,mitioglu2016magnetoexcitons}. Intermediate alloy compositions follow the same trend, consistent with the evolution of vibrational properties upon chalcogen substitution. To rationalize the composition dependence of the average phonon energy extracted from the PL fitting, we employ a simple mass--spring description, in which the characteristic vibrational frequency scales as $\omega \propto \sqrt{k/\mu}$ \cite{ashcroft2021solid}. Here, $k$ represents an effective restoring force associated with the Mo--chalcogen bonds, while $\mu$ is the reduced mass of the vibrating atomic pair. For the alloy $\mathrm{MoS}_{2x}\mathrm{Se}_{2(1-x)}$, the composition-dependent chalcogen mass is approximated as $M_{\mathrm{ch}}(x)=xM_{\mathrm{S}}+(1-x)M_{\mathrm{Se}}$, leading to an effective reduced mass $\mu_{\mathrm{eff}}(x)=M_{\mathrm{Mo}}M_{\mathrm{ch}}(x)/(M_{\mathrm{Mo}}+M_{\mathrm{ch}}(x))$. Within the simplest approximation, assuming that the effective force constant does not vary significantly across the alloy series~\cite{caramazza2016temperature}, the phonon energy is expected to follow a reduced-mass scaling $\langle \hbar\omega \rangle(x)=A/\sqrt{\mu_{\mathrm{eff}}(x)}$, where $A$ is a constant. As shown in Fig.~\ref{fig:phonon}(f), this single-parameter model provides an excellent description of the experimental data over the full composition range. The fitted prefactor is found to be $A = (111.8 \pm 0.6)\,\mathrm{meV}\sqrt{\mathrm{amu}}$. The agreement indicates that the monotonic increase of $\langle \hbar\omega \rangle$ from MoSe$_2$ to MoS$_2$ is primarily governed by the decrease in the effective reduced mass of the Mo--chalcogen vibration. Within the experimental uncertainty, no additional composition dependence of the effective force constant is required to describe the observed trend.\\
\indent Having established the composition-dependent evolution of the electronic structure and vibrational properties, we now examine the impact of alloying on exciton circular polarization. Fig.~S5 in the Supplementary Material shows PL spectra measured at 78 K under circularly polarized excitation for the full $\mathrm{MoS}_{2x}\mathrm{Se}_{2(1-x)}$ alloy series. For a direct comparison between different compositions, the excitation energy was adjusted such that the detuning —defined as the energy difference between the excitation photon energy and the A-exciton emission— remained constant at 93 meV for all samples. This condition ensures a reliable comparison of the polarization response across the alloy series. Maintaining a fixed detuning is crucial, as the degree of valley polarization is known to depend sensitively on this parameter~\cite{kourmoulakis2023biaxial}. In particular, larger detuning leads to longer exciton thermalization times, allowing depolarization processes to occur before radiative recombination and thus reducing the observed polarization~\cite{d1971spin}. Importantly, our results reveal that exciton circular polarization monotonically increases from MoSe$_2$ to MoS$_2$ while keeping the detuning energy and temperature constant (for raw spectra, see Fig. S5, supplementary material). The increase of the circular polarization with sulfur content is shown in Fig.~\ref{fig:polarization}(a), where the polarization rises steadily from nearly zero in MoSe$_2$ to about 15\% in MoS$_2$ under identical detuning and temperature conditions. This behavior can be understood within the framework of bright--dark exciton mixing. A key quantity governing the depolarization process is the bright--dark exciton splitting, defined as $\Delta_{bd}=E_{\mathrm{bright}}-E_{\mathrm{dark}}$.  This quantity arises from the combined contributions of the conduction-band spin splitting (see Fig.~\ref{fig:polarization}(b) near the $K$ valley), the short-range exchange interaction, and the different binding energies of bright and dark excitons associated with their distinct effective masses~\cite{yang2020exciton}. When $\Delta_{bd}$ approaches zero, bright and dark exciton states become nearly degenerate and can be efficiently mixed through Rashba-type spin--orbit interactions during exciton energy relaxation. This mechanism provides a fast depolarization channel and leads to the nearly vanishing polarization observed in monolayer MoSe$_2$~\cite{dery2015polarization}. Although MoSe$_2$ exhibits a relatively large spin splitting of the conduction bands near the $K$ valley, the combined contributions yield a bright--dark exciton splitting of only about $-1.5$ meV~\cite{lu2020magnetic}, placing the system close to the regime of strongest Rashba-induced depolarization. In contrast, MoS$_2$ possesses a smaller conduction-band spin splitting but a substantially larger $\Delta_{bd}$ of approximately $+14$ meV~\cite{robert2020measurement}, moving the system away from the near-degenerate condition and strongly suppressing bright--dark exciton mixing.  The DFT calculations presented in Fig.~\ref{fig:polarization}(b) reveal the systematic evolution of the two lowest spin-split conduction bands near the $K$ point with alloy composition (see also Fig.~S3). While these calculations do not directly provide $\Delta_{bd}$, they demonstrate the progressive modification of the electronic structure, associated with the bright--dark exciton splitting. Notably, first-principles calculations have shown that the exchange contribution remains of comparable magnitude across the MoX$_2$ family, on the order of $\sim20$ meV~\cite{PhysRevB.93.121107}. Consequently, the systematic evolution of the conduction-band spin splitting revealed by our DFT calculations constitutes a primary factor governing the evolution of $\Delta_{bd}$ across the alloy series. Within this picture, increasing sulfur incorporation drives the alloy system away from the near-degenerate bright--dark exciton regime characteristic of MoSe$_2$, thereby weakening the Rashba-assisted depolarization channel and leading to the monotonic enhancement of the circular polarization toward MoS$_2$~\cite{yang2020exciton}.

\section{\label{Conclusions}Conclusions}
In conclusion, our results demonstrate that chalcogen alloying in monolayer MoS$_{2x}$Se$_{2(1-x)}$ provides a powerful platform for tuning the electronic, vibrational, and spin--valley properties of two-dimensional semiconductors. By systematically varying the alloy composition, we show that the electronic structure evolves in a controlled manner, exhibiting nearly Vegard-like behavior of the excitonic transitions together with a monotonic reduction of the B--A exciton splitting from MoSe$_2$ to MoS$_2$. Density functional theory calculations reproduce the observed evolution of the electronic structure and yield spin--orbit splittings in reasonable agreement with the experimentally measured B--A exciton energy separation across the alloy series. Temperature-dependent measurements further reveal a systematic increase of the average phonon energy from Se-rich to S-rich compositions, which is well described by a reduced-mass scaling model, highlighting the dominant role of lattice mass in governing the phonon energy scale. In addition, polarization-resolved spectroscopy reveals a pronounced enhancement of the exciton circular polarization with increasing sulfur content. Supported by first-principles calculations, this behavior is attributed to alloy-induced modifications of the electronic structure that alter the bright--dark exciton splitting and consequently the efficiency of Rashba-assisted valley depolarization. Together, these findings provide new insight into the interplay between electronic structure, lattice dynamics, and spin--valley physics in TMD alloys, and establish chalcogen alloying as a versatile strategy for tailoring excitonic functionalities in two-dimensional semiconductors.

\section{\label{Exper. section}Methods}

\textit{Sample fabrication:} Monolayers were obtained by mechanical exfoliation of bulk crystals. All crystals were prepared by flux zone methods (alloy compositions are accurate within a 5\% range, as determined by energy-dispersive X-ray spectroscopy-EDS and Rutherford backscattering spectrometry-RBS measurements performed by 2D Semiconductors) except the chemical vapor  transport-grown MoSSe, used in polarization-resolved experiments, using a Nitto Denko adhesive tape and transferred onto thick hBN ($\sim$150 nm) supported on SiO$_2$/Si substrates. In addition, a thin top hBN flake ($\sim$10 nm) was subsequently placed above the monolayers to achieve full encapsulation. Encapsulation between top and bottom hBN flakes significantly improves the optical quality of the monolayers~\cite{cadiz2017excitonic} by providing atomic flatness~\cite{taniguchi2007synthesis} and clean homogeneous dielectric environment~\cite{shree2021guide}. After each transfer step, the sample was annealed at 150 $^0$C for 30 min~\cite{shree2021guide}.  

\textit{Optical Spectroscopy:} Optical spectroscopy experiments were performed using a custom-built optical setup. The excitation laser was focused onto the sample through a Mitutoyo 50x objective (NA = 0.42), producing a spot size of approximately $\sim$1 µm. The sample was mounted inside a continuous-flow cryostat (ST500, Janis), allowing temperature control during the measurements. The position of the sample was adjusted using a three-axis mechanical translation stage. The emitted signal was collected in back-reflection geometry, dispersed by a monochromator, and detected using a charge-coupled device (CCD) camera. For polarization measurements, the sample was excited with circularly polarized light, while the signal was analyzed by a combination of a liquid crystal variable retarder (LCVR) and a polarizer before the spectrometer. To ensure a consistent comparison across the alloy series, the excitation energy was adjusted such that the detuning between the excitation photon energy and the A-exciton emission remained constant at 93 meV for all compositions. 
The degree of circular polarization, which reflects the spin--valley polarization of neutral excitons, was calculated as 
\begin{equation}
P_c = \frac{I_{+} - I_{-}}{I_{+} + I_{-}}
\label{eq:polarization}
\end{equation}
where I$_+$ and I$_-$ correspond to the intensities of the $\sigma$$^+$ and $\sigma$$^-$ emission components~\cite{xiao2012coupled, cao2012valley}, respectively.

\textit{Calculations:} First-principles calculations were performed within density functional theory (DFT) using the VASP package~\cite{kresse1993ab,kresse1996efficient}. The exchange–correlation effects were treated within the generalized gradient approximation in the Perdew–Burke–Ernzerhof (PBE) form~\cite{perdew1996generalized}, while van der Waals interactions were accounted for using the DFT-D3 method of Grimme with zero damping~\cite{grimme2010consistent}. The interaction between valence and core electrons was described using the projector augmented-wave (PAW) method~\cite{blochl1994projector,kresse1999ultrasoft}. A plane-wave cutoff energy of 500 eV and a $4\times4\times1$ Monkhorst–Pack $k$-point mesh were employed, and were verified to yield converged total energies. MoS$_{2x}$Se$_{2(1-x)}$ alloys were modeled using the special quasi-random structure (SQS) approach~\cite{lebeda2026simplysqs,van2013efficient,van2002alloy}, ensuring a representative description of substitutional disorder at each composition. The SQS method was applied to $5\times5\times1$ supercells for two compositions, $x = 0.3$ and $x = 0.7$, corresponding to different sulfur concentrations on the chalcogen sublattice. The resulting two-dimensional alloy structures were relaxed in two steps. First, a \textit{2H} bulk supercell of each composition was optimized  to determine the in-plane lattice parameter $a$. Second, monolayers were constructed from the optimized bulk structures, and only atomic positions were relaxed. For the monolayer calculations, a vacuum spacing of 20~\AA\ was introduced to eliminate spurious interactions between periodic images. The relaxations in both steps were performed using the conjugate-gradient algorithm until the residual forces were below $5\times10^{-3}$ eV/\AA. A Gaussian smearing of 0.05 eV was applied. Bandstructure unfolding was performed using the VASPKIT code~\cite{wang2021vaspkit}. The unfolded spectral function was obtained by projecting the supercell wave functions onto primitive-cell Bloch states, enabling direct comparison with the band structure of the pristine system. Spin–orbit coupling was included consistently in the bandstructure calculations.

\section{\label{Acknowledgements}Acknowledgements}

This work is partially supported by ANR-23-QUAC-0004.
I. C. G. and A. S. gratefully acknowledge the computational resources provided
by the CALMIP initiative (project P0812) and CINES, IDRIS, and TGCC, which were granted by GENCI
through allocation 2025-A0180906649.
K. W. and T. T. acknowledge support from the CREST (JPMJCR24A5), JST and World Premier International Research Center Initiative (WPI), MEXT, Japan. Z. S. was supported by ERC-CZ program (project LL2101) from Ministry of Education Youth and Sports (MEYS).
E. K., K. M., D. K., E. S., G. K., X. M., and I.P. acknowledge support by the EU-funded
DYNASTY Project, ID: 101079179, under the Horizon Europe framework
programme. This work was also partly supported by the Agence Nationale de la Recherche under the program ESR/EquipEx+ (Grant No. ANR-21-ESRE-0025) and the France 2030 government investment plan managed by the French National Research Agency
under Grant Reference No. PEPR SPIN ANR-22-EXSP0007 (SPINMAT).

\bibliography{references,references_DFT}
\end{document}